# General Solutions of Cosmological Equations with Inequalities for Perfect Fluid and Scalar Field

## A. Das[1], N. Tariq[2], R.W.M. Woodside[3]

________________________________________________________________

Spherically symmetric cosmological equations in the usual FLRW coordinates are explored, with two different sources. The first couples a perfect fluid with a quintessence scalar field and the second couples the perfect fluid with a tachyonic scalar field. In *both* cases, in the inflationary regime, the scale factor $a(t)$ satisfies $a(t) > 0, \dot{a}(t) > 0, \ddot{a}(t) > 0$. *Both* sources in the matter phase yield $a(t) > 0, \dot{a}(t) > 0$. In *both* cases and *each* phase, the *general solutions* of the differential equations together with the algebraic and differential inequalities are obtained. As special cases, exponential, hyperbolic, and power law inflation, as well as power law expansion for the matter phase are all derived from the general solutions. With recent data on baryonic matter, Hubble parameter, and deceleration parameter, the relative percentages of baryonic matter, dark matter and dark energy are computed. The quintessence model yields: 4% baryonic matter, 18% dark matter, and 78% dark energy. The tachyonic model gives: 4% baryonic matter, 36% dark matter, and 60% dark energy.

________________________________________________________________

[1] Department of Mathematics, Simon Fraser University, Burnaby, B.C., Canada.
E-mail: das@sfu.ca

[2] Department of Mathematics, Simon Fraser University, Burnaby, B.C., Canada.
E-mail: ntariq@sfu.ca

[3] Physics Dept., University College of the Fraser Valley, 33844 King Rd., Abbotsford, B.C., Canada, V2S 7M9. E-mail: rob.woodside@ucfv.ca

**1. INTRODUCTION:**

It is well known that the FLRW metric is the most suitable to describe the evolution of the universe [1-4]. A perfect fluid, or a superposition of them, as well as scalar fields [5-6], supplies the algebraic structure of the energy-momentum-stress tensor. The acceleration of the inflationary phase and that of the present epoch of the matter phase violates energy conditions [7]. A perfect fluid coupled with a scalar field can explain the observational data to a great extent. In the literature, two distinct scalar fields, namely, quintessence and tachyonic fields have been considered [7-22]. In this paper, we investigate these two scalar fields separately in both inflationary and matter phases. Mathematically, these explorations involve coupled, ordinary differential equations together with algebraic and differential inequalities. We arrive at general solutions of the relevant mathematical problems with the help of *slack functions*. To elaborate the technique of a slack function, consider the toy model of a linear inhomogeneous partial differential equation with an algebraic inequality:

$$W_{,x} + W_{,t} = [h(x+t)]^2, \qquad (x,t) \in \mathbb{R}^2 ; \qquad (1.1)$$

$$W(x,t) > 0. \qquad (1.2)$$

The general solution of (1.1) is given by:

$$W(x,t) = F(x-t) + \frac{1}{2}\int_{x_o+t_o}^{x+t} [h(u)]^2 du .$$

Here, $F$ is an *arbitrary* differentiable function. However, the general solution to (1.1) together with inequality (1.2) is furnished by:

$$W(x,t) = \exp[f(x-t)] + \frac{1}{2}\int_{x_o+t_o}^{x+t} [h(u)]^2 du , \qquad (1.3)$$

with the tacit assumption $x_o + t_o < x + t$. The arbitrary function $f$ in equation (1.3) facilitates the inequality (1.2) and is thus called the slack function.

Cosmological equations in this work deal with the expansion function $a(t)$, a scalar field $\varphi(t)$ in the quintessence model or $\Phi(t)$ in the tachyonic model, and a potential function $\tilde{V}(t) := V(\varphi(t))$ or $\tilde{V}(t) := V(\Phi(t))$. Three inequalities $\tilde{V}(t) > 0$, $\dot{\varphi}(t) > 0$, and $\dot{a}(t) > 0$, can be solved with slack functions such as $\exp[f(t)] = \tilde{V}(t)$. Unfortunately, such a slack function for $\tilde{V}$ leaves an intractable mess for the differential equations. Similarly, first using a slack function for $\dot{\varphi}$ renders the differential equations extremely difficult to solve. Remarkably, first applying the slack function to $\dot{a}$ leaves a tractable set of differential equations. Following this approach, our work provides *general solutions* of the cosmological equations containing scalar fields obeying inequalities in inflationary and matter phases.

We do not completely ignore the experimental aspects of cosmology. We have compared current observational data for values of baryonic density, Hubble parameter, and the deceleration parameter with the predictions of the scalar field cosmological solutions presented here. We conclude that for the quintessence field and an input of baryonic matter at 4%, the resulting dark matter is 18%, and the resulting dark energy is 78%. However, the tachyonic field with an input of 4% baryonic matter, predicts 36% dark matter and 60% dark energy.

## 2. PERFECT FLUID, QUINTESSENCE MODEL

The following FLRW coordinates will be employed throughout the paper:

$$ds^2 = -dt^2 + a^2(t)[\frac{dr^2}{1-kr^2} + r^2 d\theta^2 + r^2 \sin^2\theta d\phi^2],$$

$$k = 0, \pm 1, \qquad 0 < 1 - kr^2, \qquad a(t) > 0, \qquad t > 0. \tag{2.1}$$

The energy-momentum-stress tensor is given by:

$$T^{\mu}{}_{\nu} = (\rho + p)u^{\mu}u_{\nu} + p\delta^{\mu}{}_{\nu} + \varphi^{;\mu}\varphi_{,\nu} - \frac{1}{2}\varphi^{;\alpha}\varphi_{,\alpha}\delta^{\mu}{}_{\nu} + V(\varphi)\delta^{\mu}{}_{\nu} \;. \tag{2.2}$$

Here $\rho$, $p$, and $u^{\mu}$ are the fluid's energy density, pressure, and velocity respectively. The quintessence field is denoted by $\varphi$ with potential $V(\varphi)$. Spherical symmetry requires $\rho$, $p$, and $\varphi$ are functions of $r$ and $t$ only. However the isotropic cosmology (2.1) dictates that

$$u^r \equiv u^\theta \equiv u^\phi \equiv 0, \qquad u^t \equiv 1, \tag{2.3}$$

and that $\rho$, $p$, and $\varphi$ are functions of $t$ only.

Einstein's field equations:

$$\mathrm{E}^{\mu}{}_{\nu} := G^{\mu}{}_{\nu} - 8\pi T^{\mu}{}_{\nu} = 0 \tag{2.4}$$

for the metric (2.1) and the energy-momentum-stress tensor (2.2) are summarized with the dependence on $t$ suppressed and the dot denoting a derivative with respect to $t$ as in [7,22].

$$\mathrm{E}^{t}{}_{t} := -3\left[\frac{\dot{a}^2 + k}{a^2}\right] + 8\pi[\rho + V(\varphi) + \frac{1}{2}\dot{\varphi}^2] = 0, \tag{2.5a}$$

$$-\mathrm{E}^{r}{}_{r} := 2\frac{\ddot{a}}{a} + \left[\frac{\dot{a}^2 + k}{a^2}\right] + 8\pi[p - V(\varphi) + \frac{1}{2}\dot{\varphi}^2] = 0, \tag{2.5b}$$

$$\mathrm{E}^{\theta}{}_{\theta} \equiv \mathrm{E}^{\phi}{}_{\phi} \equiv \mathrm{E}^{r}{}_{r}. \tag{2.5c}$$

The equation (2.5a) is the Friedman equation. A useful linear combination of these equations is the Raychaudhuri equation:

$$\frac{a}{6}[\mathrm{E}^{t}{}_{t} - 3\mathrm{E}^{\theta}{}_{\theta}] = \ddot{a} + \frac{4\pi a}{3}[\rho + 3p + 2\dot{\varphi}^2 - 2V(\varphi)] = 0\;. \tag{2.6}$$

In case we have $\rho + 3p + 2\dot{\varphi}^2 > 0$, the corresponding term in (2.6) indicates an attractive force. Moreover, the repulsion $\ddot{a} > 0$ implies the violation of the strong energy condition. Notice that $V(\varphi) > 0$ contributes to repulsion.

The equation of motion for the quintessence field is given by:

$$\sigma := \frac{1}{a^3}[a^3\dot{\varphi}]^{\cdot} + V'(\varphi) = 0, \tag{2.7}$$

where the prime denotes a derivative with respect to $\varphi$. Furthermore, the fluid continuity and streamline motions are dictated by the well-known equation

$$\dot{\rho} + 3[\ln a]^{\cdot}[\rho + p] = 0. \tag{2.8}$$

## 2.1 QUINTESSENCE INFLATIONARY PHASE

We discuss here a slack function approach to the inflationary problem in quintessence cosmology. In this phase, it is usually assumed that any fluid contribution is negligible. Moreover; the expansion factor must satisfy the following strict inequalities:

$$a > 0, \tag{2.1.1a}$$

$$\dot{a} > 0, \tag{2.1.1b}$$

$$\ddot{a} > 0. \tag{2.1.1c}$$

We also assume that

$$\dot{\varphi} > 0 \tag{2.1.2}$$

as commonly done in inflation scenarios.

The field equations (2.5) imply for $\rho = p = 0$ that:

$$\tilde{V}(t) := V(\varphi(t)) = \frac{1}{8\pi}[\frac{\ddot{a}}{a} + 2(\frac{\dot{a}^2 + k}{a^2})], \tag{2.1.3a}$$

$$[\dot{\varphi}(t)]^2 = \frac{1}{4\pi}[\frac{\dot{a}^2 + k}{a^2} - \frac{\ddot{a}}{a}]. \tag{2.1.3b}$$

Other field equations follow from the algebraic and differential identities. There is an additional inequality that follows from the last two equations, namely,

$$0 < [\dot{\varphi}(t)]^2 < \tilde{V}(t). \tag{2.1.5}$$

The system of two differential equations (2.1.3) involves three unknown functions, $a$, $\varphi$, and $\tilde{V}$. Thus the system is underdetermined and we can prescribe *one* unknown function. However, we have to satisfy five strict inequalities (2.1.1), (2.1.2) and (2.1.5). We can furnish the general solution of the system of differential equations (2.1.3) and strict inequalities. The answer is summarized as follows:

$$a(t) = (c_2)^2 + (c_1)^2 t + \varepsilon \int\limits_0^t [\int\limits_0^\tau \exp f(u) du] d\tau, \tag{2.1.6a}$$

$$8\pi \tilde{V}(t) = \frac{1}{a^2(t)} \{ \varepsilon a(t) \exp[f(t)] + 2[k + (c_1^2 + \varepsilon \int\limits_0^t \exp[f(\tau)] d\tau)^2] \} > 0, \tag{2.1.6b}$$

$$\sqrt{4\pi}[\varphi(t) - \varphi_0] = \int\limits_0^t [a(\tau)]^{-1} \{ k + (c_1)^4 + \varepsilon[2(c_1)^2 \int\limits_0^\tau \exp f(u) du - ((c_2)^2 + (c_1)^2 \tau) \exp f(\tau)] +$$

$$+ \varepsilon^2 [(\int\limits_0^\tau \exp f(u) du)^2 - \exp f(\tau) \int\limits_0^\tau (\int\limits_0^u \exp f(v) dv) du)] \}^{\frac{1}{2}} d\tau. \tag{2.1.6c}$$

The above integrand in (2.1.6c) is assumed to be positive for a sufficiently small $\varepsilon > 0$ and a sufficiently large positive $(c_1)^2$. The constants $c_2, \varphi_0$ are arbitrary. The constants $0 < \varepsilon < 1$ and $(c_1)^4 \geq 1$ are otherwise arbitrary. Thus equations (2.1.6) contain *all* (infinitely many) solutions of the problem. The following relations furnish the physical meaning of the slack function:

$$\exp[f(t)] = \ln\{\frac{8\pi}{3\varepsilon}[V(\varphi) - (\dot{\varphi})^2]\} = (\varepsilon)^{-1} \ddot{a}(t). \tag{2.1.7}$$

Here, we furnish some popular inflationary modes and their corresponding slack functions.

### 2.1.1 QUINTESSENCE MODEL: EXAMPLE 1: EXPONENTIAL INFLATION

This scenario corresponds to a slack function of the form:

$$f(t) = \omega t + \ln(\omega^2 / \varepsilon), \qquad \omega > 0. \qquad (2.1.1.1)$$

Direct computations of the equations (2.1.6) with $(c_1)^2 = \omega$, $(c_2)^2 = 1$, and $k \geq 0$ yield:

$$a(t) = \exp(\omega t), \qquad (2.1.1.2a)$$

$$8\pi \tilde{V}(t) = 3\omega^2 + 2k \exp(-2\omega t), \qquad (2.1.1.2b)$$

$$\sqrt{4\pi}\,\omega[\varphi(t) - \varphi_0] = \sqrt{k}\,(1 - \exp(-\omega t)), \qquad (2.1.1.2c)$$

$$8\pi V(\varphi) = 3\omega^2 + 2[\sqrt{4\pi}\,\omega(\varphi_0 - \varphi) - \sqrt{k}\,]^2. \qquad (2.1.1.2d)$$

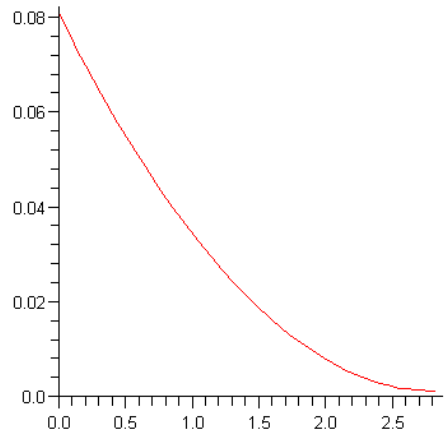

**Figure 1:** Exponential inflation in the Quintessence model. $\tilde{V}(t)$ is the ordinate and $\varphi(t)$ is the abscissa. Values of the constants for the plot: $\varphi_0 = 0$, $k = 1$, $\omega = 0.1$.

### 2.1.2 QUINTESSENCE MODEL: EXAMPLE 2: COSHYPERBOLIC INFLATION

Here the slack function is chosen to be:

$$\exp[f(t)] = (\frac{\beta^2 A}{\varepsilon})\cosh(\beta t), \quad A > 0, \quad k > (A\beta)^2, \quad \beta > 0. \tag{2.1.2.1}$$

This choice along with $c_1 = 0$ and $(c_2)^2 = A$, yields:

$$a(t) = A\cosh(\beta t), \tag{2.1.2.2a}$$

$$8\pi \tilde{V}(t) = \beta^2[1 + 2\tanh^2(\beta t)] + \frac{2k}{A^2}\cosh^{-2}(\beta t), \tag{2.1.2.2b}$$

$$\sqrt{4\pi}[\varphi(t) - \varphi_0] = \frac{1}{\beta}\sqrt{\frac{k}{A^2} - \beta^2} \, \arcsin[\tanh(\beta t)], \tag{2.1.2.2c}$$

$$8\pi V(\varphi) = 3\beta^2 + 2(\frac{k}{A^2} - \beta^2)\cos^2[\sqrt{\frac{4\pi}{\frac{k}{A^2} - \beta^2}}\beta(\varphi - \varphi_0)]. \tag{2.1.2.2d}$$

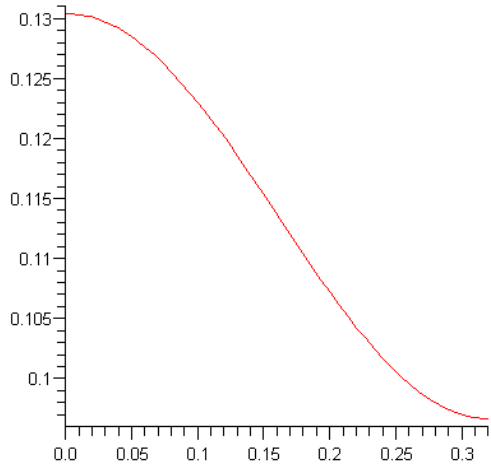

**Figure 2:** Coshyperbolic inflation in the Quintessence model. $\tilde{V}(t)$ is the ordinate and $\varphi(t)$ is the abscissa. Values of the constants for the plot: $\varphi_0 = 0$, $k = 1$, $A = \beta = 0.9$.

### 2.1.3 QUINTESSENCE MODEL: EXAMPLE 3: POWER LAW INFLATION

Here we choose:

$$\exp[f(t)] = \frac{n(n-1)}{\varepsilon}t^{n-2}, \quad n > 2, \quad k \geq 0, \quad c_1 = c_2 = 0. \tag{2.1.3.1}$$

Thus, we obtain a big bang model:

$$a(t) = t^n,$$ (2.1.3.2a)

$$8\pi\tilde{V}(t) = \frac{3n^2 - n}{t^2} + \frac{2k}{t^{2n}},$$ (2.1.3.2b)

$$\sqrt{4\pi}\varphi(t) = \frac{1}{n-1}[\sqrt{n}\ln|\sqrt{nt^{n-1}} + \sqrt{k+nt^{2n-2}}| - \sqrt{n+kt^{2-2n}}] + C.$$ (2.1.3.2c)

Here, $C$ is some constant. The last two equations define $V(\varphi)$ parametrically for $t > \varepsilon_1$, a small positive number.

## 2.2 QUINTESSENCE SCALAR FIELD IN MATTER PHASE

In the matter phase, the fluid pressure is usually assumed to be negligible compared to their energy density, allowing

$$\rho > 0, \qquad p = 0.$$ (2.2.1a)

This dust with quintessence phase is characterized by:

$$\dot{\varphi} \neq 0.$$ (2.2.1b)

The continuity equation (2.8) for the fluid flow gives a constant $M_0$ by:

$$\frac{4\pi}{3}\rho a^3 = M_0.$$ (2.2.2)

Physically, this positive constant is the total mass that is not attributable to the scalar field. The onset of the matter phase occurs at $t_2$ and the FLRW metric for $t > t_2$ provides the following field equations:

$$8\pi\tilde{V}(t) = 2[\frac{\dot{a}^2 + k}{a^2}] + \frac{\ddot{a}}{a} - \frac{3M_0}{a^3},$$ (2.2.3a)

$$4\pi[\dot{\varphi}(t)]^2 = \frac{\dot{a}^2 + k}{a^2} - \frac{\ddot{a}}{a} - \frac{3M_0}{a^3} > 0. \tag{2.2.3b}$$

The strict inequalities are assumed to be

$$a > 0, \qquad \dot{a} > 0. \tag{2.2.4}$$

Using the slack function $f(t)$, the general solution of all the field equations (2.2.3)

together with all the inequalities (2.2.1) and (2.2.4) are furnished by:

$$a(t) = \exp(c_1) + \int_{t_2}^{t} \exp[\varepsilon f(\tau)]d\tau, \tag{2.2.5a}$$

$$\rho(t) = \frac{3M_0}{4\pi}[\exp(c_1) + \int_{t_2}^{t} \exp[\varepsilon f(\tau)]d\tau]^{-3}, \tag{2.2.5b}$$

$$8\pi\tilde{V}(t) = \frac{1}{a(t)}\{2(k + \exp[2\varepsilon f(t)]) + \varepsilon \dot{f}(t)\exp[\varepsilon f(t)](\exp[c_1]\int_{t_2}^{t}\exp[\varepsilon f(\tau)]d\tau) +$$

$$-3M_0(\exp[c_1] + \int_{t_2}^{t}\exp[\varepsilon f(\tau)]d\tau)^{-1}\}, \tag{2.2.5c}$$

$$\sqrt{4\pi}[\varphi(t) - \varphi_2] = \int_{t_2}^{t}\frac{1}{a(\tau)}\{k + \exp[2\varepsilon f(\tau)] - \varepsilon \dot{f}(\tau)\exp[\varepsilon f(\tau)](\exp[c_1] + \int_{t_2}^{\tau}\exp[\varepsilon f(u)]du) +$$

$$-3M_0(\exp[c_1] + \int_{t_2}^{\tau}\exp[\varepsilon f(u)]du)^{-1}\}^{\frac{1}{2}}d\tau. \tag{2.2.5d}$$

Here, the three free parameters are $-\infty < c_1 < \infty$, $0 < \varepsilon \le 1$, and $-\infty < \varphi_2 < \infty$. The slack function $f(t)$ is of class $C^2$, but otherwise arbitrary. Its physical significance is given by the speed of expansion $\dot{a}(t)$, which is equal to $\exp[\varepsilon f(t)]$. Furthermore, if we impose the

following inequalities

$$\dot{\varphi}(t) > 0, \qquad \tilde{V}(t) > 0,$$

then $\varepsilon$ and $M_0$ must be of sufficiently small positive values.

We can obtain the Hubble parameter and deceleration parameter from equations (2.2.5) as

$$H(t) := \frac{\dot{a}(t)}{a(t)} = \exp[\varepsilon f(t)](\exp[c_1] + \int_{t_2}^{t} \exp[\varepsilon f(\tau)] d\tau)^{-1}, \qquad (2.2.6a)$$

$$q(t) := -\frac{a(t)\ddot{a}(t)}{[a(t)]^2} = -\varepsilon \dot{f}(t) \exp[-\varepsilon f(t)](\exp[c_1] + \int_{t_2}^{t} \exp[\varepsilon f(\tau)] d\tau), \qquad (2.2.6b)$$

$$\operatorname{sgn}[q(t)] = -\operatorname{sgn}[\dot{f}(t)]. \qquad (2.2.6c)$$

## 2.2.1 QUINTESSENCE MATTER PHASE: EXAMPLE 4: POWER LAW EXPANSION

Consider the slack function given by $\exp[\varepsilon f(t)] = \varepsilon n t^{n-1}$. This choice together with $k = 1$, and $\exp[c_1] = 1 + \varepsilon(t_2)^n$, provides the following solution:

$$a(t) = 1 + \varepsilon t^n, \qquad (2.2.1.1a)$$

$$8\pi \tilde{V}(t) = [1 + \varepsilon t^n]^{-2}[2 + \varepsilon n(n-1)t^{n-2} + \varepsilon^2 n(3n-1)t^{2n-2} - 3M_0(1 + \varepsilon t^n)^{-1}], \qquad (2.2.1.1b)$$

$$\sqrt{4\pi}[\varphi(t) - \varphi_2] = \int_{t_2}^{t} [1 + \varepsilon \tau^n]^{-1}[1 - \varepsilon n(n-1)\tau^{n-2} + \varepsilon^2 n\tau^{2n-2} - 3M_0(1 + \varepsilon \tau^n)^{-1}]^{\frac{1}{2}} d\tau. \quad (2.2.1.1c)$$

## 2.3 OBSERVATIONAL EVIDENCE FOR QUINTESSENCE MODEL

Now we shall compare theoretical predictions with the current cosmological observations.

In terms of the Hubble parameter $H(t)$ and the deceleration parameter $q(t)$, defined in equations (2.2.5), the field equations (2.5a) and (2.6) yield at the present time $t_p$,

$$\frac{3H^2(t_p)}{8\pi} = \rho(t_p) + V[\varphi(t_p)] + \frac{1}{2}[\dot{\varphi}(t_p)]^2 - \frac{3k}{8\pi a^2(t_p)}, \qquad (2.3.1a)$$

$$H^2(t_p)[2q(t_p)-1] = \frac{k}{a^2(t_p)} + 8\pi[\frac{1}{2}\dot{\varphi}^2(t_p) - V(\varphi(t_p))] \,. \qquad (2.3.1b)$$

Current observations provide:

$$H^2(t_p) \simeq 7.3 \cdot 10^{-53} m^{-2}, \qquad (2.3.2a)$$

$$q(t_p) \simeq -0.4 \,, \qquad (2.3.2b)$$

$$\rho(t_p) \simeq 35.5 \cdot 10^{-56} m^{-2}. \qquad (2.3.2c)$$

The equations (2.5a) and (2.2.2) yield the energy conservation,

$$\frac{1}{2}\dot{a}^2 - \frac{M_{eff}}{a} = const. \,,$$

as

$$\frac{1}{2}\dot{a}^2 - \frac{\{M_0 + \frac{4\pi}{3}[V(\varphi) + \frac{1}{2}\dot{\varphi}^2]a^3\}}{a} = -\frac{k}{2}. \qquad (2.3.3)$$

By equations (2.3.1) with $k = 1 \le a^2(t_p)$, the observed values (2.3.2), the Raychaudhuri equation (2.6), and an assumption that the perfect fluid baryonic matter is 4%, we conclude from equation (2.3.3) that the dark matter from the $\dot{\varphi}$ contribution is 18% and the repulsive dark energy arising from the $V(\varphi)$ is 78%.

## 3 PERFECT FLUID - TACHYONIC MODEL

In this model the energy-momentum-stress tensor is given by:

$$T^{\mu}_{\ \nu} = (\rho + p)u^{\mu}u_{\nu} + p\delta^{\mu}_{\nu} + \frac{V(\Phi)\Phi_{,\nu}\Phi^{;\mu}}{\sqrt{1 + \Phi_{,\alpha}\Phi^{;\alpha}}} - V(\Phi)\sqrt{1 + \Phi_{,\alpha}\Phi^{;\alpha}} \ \delta^{\mu}_{\nu}, \qquad (3.1)$$

where $\Phi$ is the "tachyonic" scalar field motivated by string theory. The Einstein field equations for this system in FLRW coordinates are:

$$E^t_{\ t} = -3[\frac{\dot{a}^2 + k]}{a^2} + 8\pi\{\rho + V[\sqrt{1 - \dot{\Phi}^2} + \frac{\dot{\Phi}^2}{\sqrt{1 - \dot{\Phi}^2}}]\} = 0, \tag{3.2a}$$

$$-E^r_{\ r} = 2\frac{\ddot{a}}{a} + \frac{\dot{a}^2 + k}{a^2} + 8\pi\{p - V\sqrt{1 - \dot{\Phi}^2}\} = 0, \tag{3.2b}$$

$$E^\theta_{\ \theta} \equiv E^\phi_{\ \phi} \equiv E^r_{\ r}. \tag{3.2c}$$

A linear combination of the above equations yields the Raychaudhuri equation:

$$\ddot{a} + \frac{4\pi a}{3}\{\rho + 3p + V[\frac{\dot{\Phi}^2}{\sqrt{1 - \dot{\Phi}^2}} - 2\sqrt{1 - \dot{\Phi}^2}]\} = 0. \tag{3.3}$$

The equation of motion for the $\Phi$ field is given by

$$\frac{1}{a^3}[\frac{a^3 V(\Phi)\dot{\Phi}}{\sqrt{1 - \dot{\Phi}^2}}]^{\cdot} = -V'(\Phi)\sqrt{1 - \dot{\Phi}^2}. \tag{3.4}$$

The fluid continuity remains the same as in equation (2.8).

## 3.1 TACHYONIC INFLATIONARY PHASE

Here, as with quintessence, we assume that the fluid contribution is negligible. Thus we set $\rho = p = 0$.

Moreover, as before, the expansion factor must satisfy the strict inequalities:

$$a > 0, \tag{3.1.1a}$$

$$\dot{a} > 0, \tag{3.1.1b}$$

$$\ddot{a} > 0. \tag{3.1.1c}$$

Also we assume that the condition,

$$\dot{\Phi} > 0, \tag{3.1.2}$$

holds and that $\dot{\Phi}$ is sufficiently small to provide for an inflation of long duration.

The system of field equations (3.2) and (3.4) boil down to

$$8\pi\tilde{V}\sqrt{1-\dot{\Phi}^2} = 2\frac{\ddot{a}}{a} + \frac{\dot{a}^2+k}{a^2}, \tag{3.1.3a}$$

$$\frac{8\pi\tilde{V}}{\sqrt{1-\dot{\Phi}^2}} = 3\frac{\dot{a}^2+k}{a^2}. \tag{3.1.3b}$$

The other field equations follow from linear combinations of these or differential identities on them.

The general solutions of the above equations (3.1.3) and strict inequalities (3.1.1) can be expressed in terms of three arbitrary constants $c_1, c_2, \varepsilon > 0,$ and an arbitrary differentiable slack function $h(t).$ The solutions are summarized below as:

$$a(t) = (c_2)^2 + t\exp[c_1] + \varepsilon\int_0^t \{\int_0^\tau \exp[h(u)]du\}d\tau, \tag{3.1.4a}$$

$$\frac{8\pi}{\sqrt{3}}\tilde{V}(t) = [a(t)]^{-2}\{[k + (\exp[c_1] + \varepsilon\int_0^t \exp[h(\tau)d\tau)^2](k + [\exp[c_1] + \varepsilon\int_0^t \exp[h(\tau)d\tau]^2 +$$

$$+2a(t)\varepsilon\exp[h(t)])\}^{\frac{1}{2}}, \tag{3.1.4b}$$

$$\sqrt{\frac{3}{2}}\ [\Phi(t) - \Phi_0] = \int_0^t \{1 - \frac{\varepsilon\exp[h(\tau)]a(\tau)}{k + (\exp[c_1] + \varepsilon\int_0^\tau \exp[h(u)]du)^2}\}^{\frac{1}{2}}d\tau. \tag{3.1.4c}$$

We shall now provide some specific examples.

### 3.1.1 TACHYONIC MODEL: EXAMPLE 1: EXPONENTIAL INFLATION

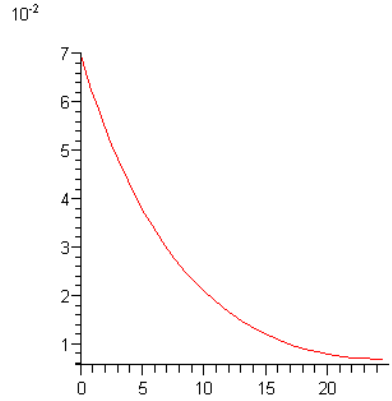

**Figure 3:** Exponential inflation in the Tachyonic model. $\tilde{V}(t)$ is the ordinate, scaled by a factor of $10^2$, and $\Phi(t)$ is the abscissa. Values of the constants for the plot: $\Phi_0 = 0$, $k = 1$, $\omega = 0.1$.

Here we choose

$$k = \varepsilon = c_2 = 1, \qquad \exp[c_1] = \omega > 0, \qquad h(t) = \omega t + 2\ln\omega, \tag{3.1.1.1}$$

to simply equation (3.1.4a) to

$$a(t) = \exp[\omega t]. \tag{3.1.1.2a}$$

Now the equations (3.1.4b) and (3,14c) yield:

$$8\pi \tilde{V}(t) = \sqrt{3(\omega^2 + k\exp[-2\omega t])(3\omega^2 + k\exp[-2\omega t])} \tag{3.1.1.2b}$$

$$\Phi(t) - \Phi_0 = \sqrt{\frac{1}{6}} \, \frac{1}{\omega} (\ln\left|\frac{\sqrt{1 + \omega^2\exp[2\omega t]} - 1}{\sqrt{1 + \omega^2\exp[2\omega t]} + 1}\right| - \ln\left|\frac{\sqrt{1 + \omega^2} - 1}{\sqrt{1 + \omega^2} + 1}\right|), \tag{3.1.1.2c}$$

$$8\pi V(\Phi) = \sqrt{3}\omega^2 \cosh[\sqrt{\frac{3}{2}}\omega(\Phi - \Phi_0) + \frac{1}{2}\ln(\frac{\sqrt{1 + \omega^2} - 1}{\sqrt{1 + \omega^2} + 1})] \times$$

$$\sqrt{2 + \cosh^2[\sqrt{\frac{3}{2}}\omega(\Phi - \Phi_0) + \frac{1}{2}\ln(\frac{\sqrt{1 + \omega^2} - 1}{\sqrt{1 + \omega^2} + 1})]}. \tag{3.1.1.2d}$$

### 3.1.2 TACHYONIC MODEL: EXAMPLE 2: COSHYPERBOLIC INFLATION

For this we choose

$$k = \varepsilon = 1, \qquad (c_2)^2 = \frac{1}{\beta} > 0, \qquad \kappa_1 := \exp[c_1] = 0, \qquad h(t) = \ln(\beta \cosh \beta t), \qquad (3.1.2.1)$$

reducing equation (3.1.4a) to

$$a(t) = \frac{\cosh \beta t}{\beta}. \qquad (3.1.2.2a)$$

Computations from equations (3.14b) and (3.1.4c) then give:

$$\Phi(t) = \Phi_0, \qquad (3.1.2.2b)$$

$$8\pi V(\Phi) = 3\beta^2. \qquad (3.1.2.2c)$$

Here $8\pi V(\Phi)$ has become a cosmological constant!

### 3.1.3 TACHYONIC MODEL: EXAMPLE 3: POWER LAW INFLATION

The choice

$$k \geq 0, \qquad \varepsilon = 1, \qquad c_2 = 0, \qquad \kappa_1 := \exp[c_1] = 0, \qquad h(t) = \ln[n(n-1)t^{n-2}], \qquad n \geq 2, \qquad (3.1.3.1)$$

simplifies equation (3.1.4a) to the big bang model:

$$a(t) = t^n, \qquad (3.1.3.2a)$$

and equations (3.1.4b) and (3.1.4c) yield:

$$8\pi \tilde{V}(t) = \sqrt{3}\, t^{-2n} \{ [k + n^2 t^{2n-2}][k + n^2 t^{2n-2} + 2n(n-1)t^{2n-2}] \}^{\frac{1}{2}}, \qquad (3.1.3.2b)$$

$$\Phi(t) - \Phi_0 = \sqrt{\frac{2}{3}} \int_0^t \left[ \frac{k + n\tau^{2n-2}}{k + n^2 \tau^{2n-2}} \right]^{\frac{1}{2}} d\tau. \qquad (3.1.3.2c)$$

### 3.2 TACHYONIC SCALAR FIELD IN MATTER PHASE

In this phase, we adopt the same dust assumption as in equation (2.2.1a) so that the continuity equation (2.8) gives the same constant $M_0$ by equation (2.2.2) for the tachyonic case as well.

The field equations (3.2) in the time interval $(t_2, t)$, with $0 < t_2$, go over into:

$$[8\pi\tilde{V}(t)]^2 = [8\pi V(\Phi(t))]^2 = 3[\frac{\dot{a}^2 + k}{a^2} - \frac{2M_0}{a^3}][\frac{2\ddot{a}}{a} + \frac{\dot{a}^2 + k}{a^2}], \qquad (3.2.1a)$$

$$\frac{3}{2}[(\dot{\Phi}(t))]^2 = [\frac{\dot{a}^2 + k}{a^2} - \frac{\ddot{a}}{a} - \frac{3M_0}{a^3}]/[\frac{\dot{a}^2 + k}{a^2} - \frac{2M_0}{a^3}], \qquad (3.2.1b)$$

$$\ddot{a}(t) = -\frac{4\pi}{3}[\rho(t) + \frac{\tilde{V}(t)\dot{\Phi}^2}{\sqrt{1-\dot{\Phi}^2}} - 2\tilde{V}(t)\sqrt{1-\dot{\Phi}^2}]a(t). \qquad (3.2.1c)$$

The strict inequalities are assumed to be

$$a(t) > 0, \qquad \dot{a}(t) > 0, \qquad \dot{\Phi}(t) > 0, \qquad \tilde{V}(t) > 0. \qquad (3.2.2)$$

The general solutions of the field equations (3.2.1) together with the inequalities (3.2.2) are furnished by:

$$a(t) = \exp[c_1] + \int_{t_2}^{t} \exp[\varepsilon h(\tau)]d\tau, \qquad (3.2.3a)$$

$$\frac{8\pi}{\sqrt{3}}\tilde{V}(t) = [a(t)]^{-2}\{[k + \exp[2\varepsilon h(t)] - \frac{2M_0}{a(t)}][k + \exp[2\varepsilon h(t)] + 2\varepsilon\dot{h}(t)\exp[\varepsilon h(t)]a(t)]\}^{\frac{1}{2}},$$

$$(3.2.3b)$$

$$\sqrt{\frac{3}{2}}(\Phi(t) - \Phi_2) = \int_{t_2}^{t}\{\frac{k + \exp[2\varepsilon h(\tau)] - \varepsilon\dot{h}(\tau)\exp[\varepsilon h(\tau)]a(\tau) - 3M_0[a(\tau)]^{-1}}{k + \exp[2\varepsilon h(\tau)] - 2M_0[a(\tau)]^{-1}}\}^{\frac{1}{2}}d\tau. \quad (3.2.3c)$$

Here, $-\infty < c_1 < \infty$, $0 < \varepsilon \leq 1$, $M_0$ has a small positive value and the slack function $h(t)$ is twice differentiable, but otherwise arbitrary. Note that equations (3.2.3) contain *all* solutions of the matter domain.

Now we shall provide a specific example, namely the *power law*. To do so we choose

$$\varepsilon = 1, \qquad c_1 = n\ln(t_2), \qquad n \geq 1, \qquad (3.2.4a)$$

and the slack function to be

$$h(t) = \ln(nt^{n-1}). \qquad (3.2.4b)$$

By the general solution (3.2.3) and these choices, we obtain:

$$a(t) = t^n, \qquad (3.2.5a)$$

$$\frac{8\pi}{\sqrt{3}}\tilde{V}(t) = t^{-2n}\{[k + n^2 t^{2n-2} - 2M_0 t^{-n}][k + n(3n-2)t^{2n-2}]\}^{\frac{1}{2}}, \qquad (3.2.5b)$$

$$\sqrt{\frac{3}{2}}(\Phi(t) - \Phi_2) = \int_{t_2}^{t}[\frac{k\tau^n + n\tau^{3n-2} - 3M_0}{k\tau^n + n^2\tau^{3n-2} - 2M_0}]^{\frac{1}{2}}d\tau. \qquad (3.2.5c)$$

We do assume that

$$0 < 3M_0 < k(t_2)^n + n(t_2)^{3n-2}. \qquad (3.2.6)$$

### 3.3 OBSERVATIONAL EVIDENCE FOR TACHYONIC MODEL

Now, we shall compare theoretical predictions with the observational evidence. Putting the present Hubble parameter $H(t_p)$ and the deceleration parameter $q(t_p)$ into the field equations (3.1.3) yields the following:

$$\frac{3}{8\pi}H^2(t_p) = \rho(t_p) + \tilde{V}(t_p)[\sqrt{1 - \dot{\Phi}^2} + \frac{\dot{\Phi}^2}{\sqrt{1 - \dot{\Phi}^2}}], \qquad (3.3.1a)$$

$$H^2(t_p)[2q(t_p) - 1] = \frac{k}{a^2(t_p)} - 8\pi\tilde{V}(t_p)\sqrt{1 - \dot{\Phi}^2}. \qquad (3.3.1b)$$

The energy conservation equation that follows from equation (3.2a), is given by

$$\frac{1}{2}(\dot{a})^2 - \frac{1}{a}\{M_0 + \frac{4\pi}{3}V(\Phi)[\sqrt{1 - \dot{\Phi}^2} + \frac{\dot{\Phi}^2}{\sqrt{1 - \dot{\Phi}^2}}]a^3\} = -\frac{k}{2}. \qquad (3.3.2)$$

With the assumption that the baryonic matter is 4%, the current observations (2.3.2) with $k << a^2(t_p)$, yield from equations (3.3.1) and (3.2a) that the dark matter, from $\dot{\Phi}^2$, is 36%, and the dark energy, from $V(\Phi)$, is 60%.

## 4 ACKNOWLEDGEMENTS

One of us (A. D.) had informal discussions with Drs. Andrew DeBenedictis and Steve Kloster. R.W.M.W. would like to thank George McGuire for his assistance in the use of Maple and the research office at U.C.F.V. for the sabbatical that permitted this work.

## REFERENCES

[1] Hawking, S.W., and Ellis, G.F.R. (1973). *The Large Scale Structure of Space-time*, Cambridge University Press, Cambridge, U.K.

[2] Misner, C., Thorne, K.S., and Wheeler, J.A. (1973). *Gravitation*, W.H. Freeman, San Francisco, U.S.A.

[3] Weinberg, S. (1972). *Gravitation and Cosmology*, John Wiley & Sons Inc., New York, U.S.A.

[4] Stephani, H. (1982). *General Relativity: An Introduction to the Theory of the Gravitational Field*, Cambridge University Press, Cambridge, U.K.

[5] Das, A., and Agrawal, P. (1974). *Gen. Rel. Grav.*, **5**, 359.

[6] Gegenberg, J.D., and Das, A. (1985). *Phys. Lett.* **112A**, 427.

[7] Gibbons, G.W. (2003). hep-th /0301117.

[8] Chimento, L.P., and Jakubi, A. (1999). *Phys. Rev. D.* **60**, 103501.

[9] Mak, M.K., and Harko, T.(2002). *Int. J. Mod. Phys.* **11**, 1389.

[10] Majumdur, A., Panda, S., and Pérez-Lorenzana, A. (2001). *Nucl. Phys. B* **614**, 101.

[11] Sen, A. (2002). *JHEP* **0207**, 065.

[12] Kofman, L., and Linde, A.(2002). *JHEP* **0207**, 005.

[13] Tia, Y.S., Cai,. R.G., Zhang, X., and Zhong, X.Z. (2002). *Phys. Rev. D* **66**, 121301.

[14] Frolov, A., Kofman, L., and Starobinsky, A. (2002). *Phys. Lett. B* **545**, 8.

[15] Fairbairn, M., and Tytgat, M.H.G. (2002). *Phys. Lett. B* **546**, 1.

[16] Padmanavan, T. (2002). *Phys. Rev. D* **66**, 021301.

[17] Padmanavan, T., and Roy Choudhury, T. (2002). *Phys. Rev. D* **66**, 081301.

[18] Bagla, J.S., Jassal, H.K., and Padmanavan, T. (2003). *Phys. Rev. D* **67**, 063504.

[19] Kim, C. Kim, H.B., and Kim, Y. (2003). *Phys. Lett. B* **552**, 111.

[20] Kar, S. (2002). hep-th/ 0210108.

[21] Garousi, M.R., Sami, M., and Tsujikawa, S. (2004). *Phys. Rev. D* **70**, 043536.

[22] DeBenedictis, A., Das, A., and Kloster, S. (2004). *Gen. Rel. Grav.* **36**, 2481.